\documentclass[a4paper,fleqn,usenatbib]{mnras}
\usepackage{newtxtext,newtxmath}

\usepackage[T1]{fontenc}
\usepackage{ae,aecompl}

\usepackage{graphicx}	
\usepackage{amsmath}	
\usepackage{amssymb}	

\title[Pressure and entropy of heliosheath electrons]{Concerning pressure and entropy of shock-accelerated heliosheath electrons}

\author[H.~J. Fahr and R. Dutta-Roy]{
Hans J. Fahr$^{1}$\thanks{E-mail: hfahr@astro.uni-bonn.de},
Robindro Dutta-Roy$^{1}$
\\
$^{1}$Argelander-Institut f\"ur Astronomie der Universit\"at Bonn, Auf dem H\"ugel 71, 53121 Bonn, Germany
}

\date{Accepted XXX. Received YYY; in original form ZZZ}

\pubyear{2018}

\begin{document}
\label{firstpage}
\pagerange{\pageref{firstpage}--\pageref{lastpage}}
\maketitle

\begin{abstract}
We study the
behaviour of shocked wind-electrons leaving wind-driving stars after
undergoing the outer wind termination shock.
As an example, we describe
the evolution of the keV-energetic electron distribution
function downstream of the heliospheric termination shock.
We start from a kinetic transport equation in the bulk frame of
the heliosheath plasma flow taking into account shock-induced electron
injection, convective changes, cooling processes, and whistler wave-induced
energy diffusion. From this equation we proceed to an associated
pressure moment of the electron distribution function arriving at a
corresponding pressure transport equation which describes the
evolution of the electron pressure in the bulk frame of the plasma
along the plasma flow lines. We assume that the local
distribution function, in view of the prevailing non-LTE conditions, is
represented by a local kappa function with local kappa
parameters that vary with the streamline coordinate $s$ downstream of
the solar wind termination shock.
We obtain the solution for the electron
pressure as a function of the streamline coordinate $s$
from the pressure transport equation and
demonstrate that, connected with this pressure, one obtains an expression
for the entropy of the electron fluid which can also be
derived as a streamline function. We show
that the heliosheath electron fluid can essentially be characterized as an
isobaric and isentropic flow. These results allow to generally conclude
that astrotail plasma flows are characterized as such flows.
\end{abstract}


\section{Introduction}
It is well known that downstream of the solar wind termination shock
energetic electrons in the range between 40 and 70~keV were observed by
Voyager-2 \citep{krimigis2015,deckeretal2015}.
These low- and
middle-energetic electrons in these distant space plasma regions up to now
play the role of unclassified and unpredicted particle species. 
Concerning theoretical studies, it is usually assumed at shocks like the solar
wind termination shock that these light species, i.e.\ the electrons, at the
transition from the upstream to the downstream side of the shock, strictly
follow by velocity the Rankine-Hugoniot relations valid for the main
momentum carrier, i.e.\ the ions. This means, it is generally assumed that the
electrons react like the ions concerning their moment properties, i.e.\
attaining downstream of the shock densities, bulk velocities, and
temperatures identical to those of the ions. In fact, electrons and ions at
very localized shock structures cannot be taken as strongly coupled to
each other, one species kinetically strongly bound to the other, bound
to each other like electron and proton in the form of a neutral H-atom.
Actually, the locally very strong, shock-induced electric fields lead to the
phenomenon called "spontaneous demagnetization"
\citep[see][]{lembegeetal2003}, and due
to this fact electrons in first order only react to the strong shock-induced
electric fields essentially not recognizing Lorentz forces. This pinpoints
the special role of electrons at shock passages which had been emphasized
already in many papers of the recent past like e.g.\ by
\citet{chalovfahr2013}, \citet{fahretal2014}, and \citet{fahretal2015}.
The background of this phenomenon has been clearly pointed out in a recent paper by
\citet{fahrverscharen2016}
showing that, as a consequence of a piece-wise
"de-magnetization", electrons passing over the electric shock structure gain
high overshoot velocities with energies of the order of several keV. In
the following part of this paper, we start from these keV-energetic,
shock-induced electrons and follow their kinetic fate in configuration and
velocity space at their convection downstream from the termination shock.

\section{Study of the distribution function along heliosheath flow lines}
We first come back to the kinetic phasespace transport equation which we
have already developed and used in a very similar application, namely to ions
leaving the shock in \citet{fahretal2016}.
We analogously here assume that
the locally prevailing electron distribution function $f_{e}$, shown
to be initially given by a kappa function with parameters $\Theta =\Theta_{0}$
and $\kappa =\kappa_{0}$
\citep[see e.g.][]{fahrverscharen2016,fahretal2017},
also further downstream of the termination shock can at least be
expected as a non-equilibrium kappa-like function $f_{e}^{\kappa}(\kappa,\Theta )$,
characterized, however, by locally varying kappa parameters $\kappa =\kappa(s)$
and $\Theta =\Theta(s)$ varying with the streamline coordinate $s$
along the streamlines given by:
\begin{equation}
\begin{split}
\label{eq0}
f_{e}^{\kappa} &=
f_{e}^{\kappa}(\kappa(s),\Theta(s),v) \\
&= \frac{n_{e}}{(\pi\kappa(s)\Theta^{2}(s))^{3/2}}\frac{\Gamma(\kappa(s)+1)}{\Gamma(\kappa(s)-1/2)}
[1+\frac{v^{2}}{\kappa(s)\Theta^{2}(s)}]^{-(\kappa(s)+1)}
\end{split}
\end{equation}
where $n_{e}$ is the electron density on the downstream side of the
shock, and $\kappa(s)$ and $\Theta(s)$ denote the local kappa parameters.
The function $\Gamma(x)$ is the well known mathematical Gamma function.

In fact, we are refering here to electrons that originate from upstream
thermal solar wind electrons after they have undergone the differential
acceleration due to the action of the shock-electric field. The basic
theoretical description of that process has been given in our papers
\citep{fahretal2015,fahrverscharen2016}.
Here we show
within a multifluid concept that electrons react to the shock-electric
field in a very specific way leading to downstream suprathermal,
non-equilibrium electrons. The distribution function of these latter
electrons can be well described by a Kappa-function that even extends its
high velocity wing up to energies of 30~keV. This is shown in a more recent
paper by
\citet{fahretal2017}
and in fact explains the spectral electron
fluxes measured by Voyager-1 between 40--70~keV near the termination shock
region.

The resulting distribution function is determined by
the following kinetic transport equation
\begin{equation}
\label{eq1}
\frac{\partial f_{e}^{\kappa}}{\partial t}=-U\frac{df_{e}^{\kappa}}{ds}
+\left\vert \frac{\partial f_{e}^{\kappa}}{\partial t}\right\vert_{mag}+
\frac{1}{v^{2}}\frac{\partial}{\partial v}(v^{2}D_{vv}\frac{\partial
f_{e}^{\kappa}}{\partial v})
\end{equation}
where the term on the left side denotes the local time derivative of
$f_{e}^{\kappa}$, the first term on the right side denotes the convective
derivative of $f_{e}^{\kappa}$, the second term the so-called magnetic
cooling connected with the conservation of the magnetic particle moment, and
the third term describes the effect of a wave-induced velocity-space
diffusion.
$U$ is the plasma bulk velocity.
For more explanations of the explicit form of these terms the
reader should also look into the paper by \citet{fahretal2016}.

The processes taken into account in Eq.~(\ref{eq1}) are convective changes of
the electron fluid moving with the bulk speed of the plasma flow, magnetic
cooling, and energy diffusion by interaction with whistler wave turbulences.
'Magnetic cooling' we call the effect connected with the tendency of the
electrons to keep their magnetic moment constant at the convection downflow.
As we could show in many papers in the past
\citep{fahrsiewert2010,fahrsiewert2013,fahr2007}
this process operates on all time scales,
even if other counteracting processes are operating. It does not mean that
the magnetic moment in fact is conserved, but it means that the tendency to
conserve the magnetic moment $(KT/mB)$ has to be accordingly respected in the
transport equation.

Looking next only for stationary solutions of the problem (i.e.\ $\partial
/\partial t=0$), multiplying Eq.~(\ref{eq1}) by $(4\pi/3)mv^{2}$ and
integrating over velocity space with $v^{2}dv$ following the
procedure practised by \citet{fahretal2016}
(and futheron skipping the
subindex "$e$", since only electrons are addressed in this paper), one
obtains the following pressure-moment transport equation:
\begin{equation}
\label{eq2}
\frac{dP^{\kappa}}{ds}=\frac{4\pi m}{3U}\int^{v}v^{4}dv\cdot \lbrack
\left\vert \frac{\partial f^{\kappa}}{\partial t}\right\vert_{mag}+\frac{1}{v^{2}}
\frac{\partial}{\partial v}(v^{2}D_{vv}\frac{\partial f^{\kappa}}{\partial v})]
\end{equation}

Eq.~(\ref{eq2}) contains two undeveloped integral terms which can be
written in the form
\begin{equation}
\label{eq3}
T1=\int v^{4}dv\cdot \left\vert \frac{\partial f^{\kappa}}{\partial t}\right\vert_{mag}
\end{equation}
and
\begin{equation}
\label{eq4}
T2=\int v^{2}dv\cdot \frac{\partial}{\partial v}(v^{2}D_{vv}\frac{\partial f^{\kappa}}{\partial v})
\end{equation}

The term $T1$, describing magnetic cooling due to the conservation of the
particle's magnetic moment
\citep[see][]{fahrsiewert2013},
has the following
detailed representation:
\begin{equation}
\label{eq5}
T1=\int v^{4}dv\cdot (\frac{1}{v^{2}}\frac{\partial}{\partial v}(v^{2}\dot{v}_{m}f^{\kappa})
=\int dvv^{2}\frac{\partial}{\partial v}(v^{2}\dot{v}_{m}f^{\kappa})
\end{equation}
where the magnetic velocity drift $\dot{v}_{m}$ due to magnetic moment
conservation is given by
\citep[see][]{fahr2007,fahrfichtner2011,fahrsiewert2013}
\begin{equation}
\label{eq6}
\dot{v}_{m}=U\left\vert \frac{\partial v}{\partial s}\right\vert _{m}=
\frac{2}{3}vU\frac{1}{B}\frac{\partial B}{\partial s}
\end{equation}
with $B$ being the magnetic field magnitude.

Eq.~(\ref{eq6}) describes this magnetic moment conservation tendency by
expressing the temporal change of the velocity perpendicular to the magnetic
field due to the bulk motion in a direction into which the magnetic field
magnitude $B$ changes. Hereby $v_m$ is the pitch-angle averaged velocity
perpendicular to $B$
\citep[see][]{fahr2007}.
This leads to the following
term $T1$ in the pressure transport equation.
\begin{equation}
\label{eq7}
T1=\frac{2}{3}U\frac{1}{B}\frac{\partial B}{\partial s}\int dvv^{2}
\frac{\partial}{\partial v}(v^{3}f^{\kappa})=\frac{4}{3}U\frac{1}{B}
\frac{\partial B}{\partial s}\frac{3}{4\pi m}P^{\kappa}(s)
\end{equation}
where the streamline gradient of the magnetic field magnitude, i.e.\
$\partial B/\partial s$, can perhaps best be obtained from a
paper by \citet{suessnerney1990}
presenting analytic
solutions for the frozen-in heliosheath magnetic field.
$m$ is the electron mass.

Here in this calculation we use the assumption of a kinematically
frozen-in magnetic field, in order to be able to use the available analytic
solutions for $\vec{B}=\vec{B}(\vec{r})$. In fact the self-consistency in
the field description is only relevant in the stagnation region of the
heliosheath (close to the heliopause, where magnetic reconnection effects
come into the game), but in the main region of the heliosheath which we are
describing here the kinematical approach towards the frozen-in field
$\vec{B}$ given by \citet{suessnerney1990} is viable.

In Eq.~(\ref{eq7}), the local kappa pressure $P^{\kappa}(s)$ connected
with the local parameters $\kappa(s)$ and $\Theta(s)$ is defined by
\begin{equation}
\label{eq8}
P^{\kappa}(s)=\frac{4\pi}{3}m\int v^{4}f^{\kappa}(s,v,\kappa,\Theta)dv=
\frac{1}{2}mn\Theta^{2}(s)\frac{\kappa(s)}{\kappa(s)-3/2}
\end{equation}

Finally, the second term $T2$ has extensively been analysed in
\citet{fahretal2014,fahretal2016}
and depends on the specific form of the velocity diffusion
coefficient. For a velocity-dependence of this diffusion coefficient in the
form $D_{vv}=D_{0}\cdot v^{\alpha}$, it leads to the following expression:
\begin{equation}
\label{eq9}
T2=2D_{0}(3+\alpha)\int v^{2+\alpha}f^{\kappa}dv
\end{equation}

The velocity diffusion process for electrons is based on electron
interactions with selfconsistent whistler wave turbulences. The actual
diffusion coefficient for this process is velocity-dependent and in our
calculations further down from Eq.~(\ref{eq9}) we take it to depend on
$v^2$ (i.e. $\alpha=2$). The diffusion coefficient $D_0$ is a reference value for
the velocity $v_0 = \sqrt{2~\textrm{keV}/m_e}$ and in our calculations is taken to be
$D_0 = 10^{-9} s^{-1}$.

Adopting for the velocity-dependence of the diffusion coefficient (for the
case of electrons) as a reasonable value $\alpha =2$ \citep[see][]{kallenbachetal2005}, one is led to
\begin{equation}
\label{eq10}
T2=10D_{0}\int v^{4}f^{\kappa}dv=\frac{3}{4\pi m}10D_{0}P^{\kappa}(s)
\end{equation}

The moment transport equation derived above now attains the following
explicit form:
\begin{equation}
\begin{split}
\label{eq11}
\frac{dP^{\kappa}}{ds} &= \frac{1}{U}\frac{4\pi m}{3}[T1+T2] \\
&= \frac{1}{U}
\frac{4\pi m}{3}[\frac{4}{3}U\frac{1}{B}\frac{\partial B}{\partial s}\frac{3}
{4\pi m}P^{\kappa}(s)+\frac{3}{4\pi m}10D_{0}P^{\kappa}(s)] \\
&= \frac{4}{3}\frac{1}{B}\frac{\partial B}{\partial s}P^{\kappa}(s)+\frac{1}{U}10D_{0}P^{\kappa}(s)
\end{split}
\end{equation}

From this transport equation (Eq.~\ref{eq11}), one consequently derives the
following more condensed form for the relevant pressure transport equation
\begin{equation}
\label{eq11a}
\frac{1}{P^{\kappa}}\frac{dP^{\kappa}}{ds}=\frac{4}{3}\frac{1}{B}\frac{\partial B}{\partial s}+\frac{10D_{0}}{U}
\end{equation}
yielding $P^{\kappa}(s)$ as a function of the line element $s$,
interestingly enough without knowledge of $\kappa(s)$ and $\Theta(s)$, in
the following explicit form:
\begin{equation}
\begin{split}
\label{eq12}
P^{\kappa}(s) &= P^{\kappa}(s_{0})\cdot \exp [\int_{s_{0}}^{s}
(\frac{4}{3}\frac{1}{B}\frac{\partial B}{\partial s}+
\frac{1}{U}10D_{0})ds] \\
&= P^{\kappa}(s_{0})\cdot (\frac{B(s)}{B(s_{0})})^{4/3}
\exp[10D_{0}\int_{s_{0}}^{s}\frac{ds}{U}]
\end{split}
\end{equation}
where $P^{\kappa}(s_{0})$ denotes the initial pressure at the entrance
coordinate $s_{0}$ into the heliosheath, the place where the plasma flow
enters the heliosheath (i.e.\ at the termination shock with initial
streamline coordinate $s=s_{0}$). This initial pressure $P^{\kappa }(s_{0})$
near the nose (upwind direction) can be calculated in absolute values based
on the following quantities taken from
\citet{fahretal2017}:
$n_{e}(s_{0})=1.5 \cdot 10^{-3}{\rm cm}^{-3}$, $m\Theta^{2}(s_{0})=E_{0}=1 {\rm keV}$, $\kappa(s_{0})=7$,
and allows to calculate the absolute electron pressure
along heliosheath streamlines.

Astonishingly, it is evident from Eq.~(\ref{eq12}) that the actual
pressure $P^{\kappa}(s)$ is influenced only by:

A: the change of the magnetic field magnitude along the streamline,

and by

B: the diffusivity parameter
$d_{vv} = D_{0} \int_{s_{0}}^{s}\frac{ds}{U}$ with $D_{0}=10^{-9} {\rm s}^{-1}$.

As one can easily check, this latter factor in the upwind heliosheath
evaluates to the order of $d_{vv} \leq 10^{-2}$ \citep[see][]{chalovetal2007},
meaning that by this expression
pressure changes of about a factor of $\exp(1.0)$ are indicated. Also, the remaining first factor
in the upper relation in front of the exponential function is given by the
ratio of the field magnitudes at the streamline points $s$ and $s_{0}$,
respectively. It is fairly evident that also this factor does not describe more
essential pressure changes along the streamlines. Thus, it can be concluded
that the heliosheath electron fluid behaves essentially as an "isobaric"
medium. This means that the heliosheath streamlines
\citep[see][]{syllafichtner2015}
essentially represent electron
pressure isobars. Connected with earlier arguments given elsewhere
\citep{fahrsiewert2015}
which showed that due to the very low sonic Mach numbers
the plasma behaves incompressible, this also means that $T(s)=P(s)/(n(s)K)$ is
fairly constant along streamlines and hence streamlines in the heliosheath
characterize electron isothermals.

At a first glance, it is perhaps interesting to see that the resulting
electron pressure $P^{\kappa}(s)$, as evident from Eq.~(\ref{eq12}),
depends directly neither on $\kappa(s)$
nor on $\Theta(s)$ as could have been expected in principle from the fact
that this pressure is locally defined on the basis of these two parameters
through the expression
\citep[see][]{heerikhuisenetal2008,schereretal2018}
\begin{equation}
\label{eq13}
P^{\kappa}(s)=\frac{1}{2}mn[\Theta^{2}(s)\frac{\kappa(s)}{\kappa(s)-3/2}]
\end{equation}
with parameters $\kappa =\kappa(s)$ and $\Theta =\Theta(s)$ varying along
the streamline.

However, it must first be kept in mind that any combination of parameters
$\kappa$ and $\Theta$, solely fulfilling the following relation
\begin{equation}
\label{eq14}
\Theta^{2}(s)\frac{\kappa(s)}{\kappa(s)-3/2}=\frac{2P^{\kappa}(s)}{mn}
\end{equation}
is a permitted solution and that in this respect $P^{\kappa}(s)$,
as evident, depends directly neither on $\kappa$ nor on $\Theta$.

And second, it has to be kept in mind that the physically deeper-rooted
explanation of the above mentioned independence of the pressure $P(s)$ on
the local parameters $\kappa(s)$ and $\Theta(s)$ may, however, in truth
rather be that the above pressure transport equation could have been derived
without prespecifying at all the pressure $P=P(s)$ as a kappa pressure
$P^{\kappa}(s)$. One namely would arrive at an identical
pressure transport equation (Eq.~\ref{eq12}) without prescribing apriori anything about
the underlying distribution function.

\section{The electron entropy and a thermodynamic streamline constant for isentropic flows}
In the ongoing consideration we study the thermodynamical behaviour of the
electron population being convected with the plasma bulk flow downstream
from the termination shock. We start out from the well known thermodynamic
relation connecting changes of the internal energy $d\varepsilon$ of a plasma
volume $dV$ and the pressure with changes of the entropy $S$ in the well known
classic thermodynamical form
\begin{equation}
\label{eq15}
TdS = d\varepsilon + PdV
\end{equation}
where $T$ and $P$ denote temperature and pressure of the electron fluid,
$d\varepsilon$ is the internal energy of the volume $dV$, and
$S=S_{e}$ denotes the electron fluid entropy. Let us then first look here
onto isentropic flows, i.e.\ those flows for which in the upper relation the
change of the entropy $S$ with the streamline coordinate $s$ vanishes, i.e.\
$dS/ds=0$. This request means that along streamlines the electron fluid
behaves isentropic and purely adiabatic.

For such isentropic flows, the above relation then simply states that the
work done by the pressure at the expansion of the co-moving plasma volume
$\Delta V$ is the only reason for a reduction or an increase of the inner
electron energy $d\varepsilon$ of this volume. For that latter case, one hence
obtains the following relation between pressure and energy density
$\varepsilon$ for a co-moving plasma volume $\Delta V$ (i.e.\ one which is
flanked by adjacent streamlines) of the electron fluid at an increment $ds$
of the streamline coordinate $s$:
\begin{equation}
\label{eq16}
-p\frac{d}{ds}\Delta V=\frac{d}{ds}(\varepsilon \cdot \Delta V)
\end{equation}
where the relation between pressure and energy density $\varepsilon$ of the
electrons via moment-definitions is given by $p=\frac{4\pi}{3}\varepsilon$.
This furtheron leads to the following relation, if again here we
assume to be based on kappa functions and the corresponding kappa pressure
$P^{\kappa}(s)$ (see Eq.~\ref{eq12}):
\begin{equation}
\begin{split}
\label{eq17}
-[\frac{1}{2}mn\Theta (s)^{2}\frac{\kappa(s)}{\kappa(s)-3/2}]\cdot
\frac{d}{ds}\Delta V \\
=\frac{d}{ds}([\frac{3}{4\pi}\frac{1}{2}mn\Theta(s)^{2}
\frac{\kappa(s)}{\kappa(s)-3/2}]\cdot \Delta V)
\end{split}
\end{equation}
which furthermore, in view of the fact that the low-Mach-number heliosheath
plasma flow with Mach numbers $M_{s}\leq 0.1$ can be considered as behaving
incompressible (i.e.\ $dn/ds=0$, see argumentation given e.g.\ in
\citet{fahrsiewert2015},
then leads to
\begin{equation}
\label{eq18}
-\Theta ^{2}(s)\frac{\kappa(s)}{\kappa(s)-3/2}\frac{d}{ds}\Delta V
=\frac{d}{ds}(\frac{3}{4\pi}\Theta^{2}(s)\frac{\kappa(s)}{\kappa(s)-3/2}\Delta V)
\end{equation}
or yielding
\begin{equation}
\begin{split}
\label{eq19}
-\frac{1}{\Delta V}\frac{d}{ds}\Delta V &= \frac{1}{(1+\frac{4\pi }{3})}
\frac{d}{ds}\ln [\Theta _{0}^{2}\bar{\Theta}^{2}(s)\frac{\kappa (s)}{\kappa(s)-3/2}] \\
&= \frac{3}{3+4\pi}\frac{d}{ds}[\ln \bar{\Theta}^{2}+\ln \frac{\kappa(s)}{\kappa(s)-3/2}+\ln \Theta_{0}^{2}]
\end{split}
\end{equation}

The relation above (Eq.~\ref{eq19}) can then be put into the following form:
\begin{equation}
\label{eq20}
\frac{d}{ds}[\ln \Delta \bar{V}+\frac{3}{(3+4\pi )}\ln [\bar{\Theta}^{2}(s)
\frac{\kappa(s)}{\kappa(s)-3/2}]]=\frac{d}{ds}\eta =0
\end{equation}
where we have introduced in the formula above normalizations of the velocity
by the reference value $\Theta_{0}$ and of the volume by the reference
value $V_{0}$, respectively. Eq.~(\ref{eq20}) shows that the
quantity $\eta$ for isentropic flows represents a constant on streamlines,
meaning that for any selected streamline parameter $\Lambda$
\citep[for its definition see][]{syllafichtner2015}
it is thus given by
\begin{equation}
\label{eq21}
\eta(\Lambda)=\left \vert \ln \Delta \bar{V}+\frac{3}{(3+4\pi )}\ln
[\bar{\Theta}^{2}(s)\frac{\kappa(s)}{\kappa(s)-3/2}]\right\vert_{\Lambda}
\end{equation}
and, when taking everything to the "$e$"-th power, leading to the requirement
\begin{equation}
\begin{split}
\label{eq22}
\exp [\eta] &= \Delta \bar{V}_{s_{1}}\cdot \lbrack \bar{\Theta}^{2}(s_{1})
\frac{\kappa(s_{1})}{\kappa(s_{1})-3/2}]^{\frac{3}{3+4\pi }} \\
&= \Delta \bar{V}_{s_{2}}\cdot \lbrack \bar{\Theta}^{2}(s_{2})\frac{\kappa(s_{2})}{\kappa(s_{2})-3/2}]^{\frac{3}{3+4\pi}}
\end{split}
\end{equation}

Hereby the co-moving fluid volume $\Delta V$ is connected with the streamline
geometry of the plasma flow and can be defined by means of the particle
conservation requirement for the co-moving fluid volume for
$s_{0}<s_{1}<s_{2}$ in the following form:
\begin{equation}
\label{eq23}
n_{s_{0}}\Delta V_{s_{0}}U_{s0}=n_{s_{1}}\Delta
V_{s_{1}}U_{s_{1}}=n_{s_{2}}\Delta V_{s_{2}}U_{s_{2}}
\end{equation}

At least in that region of the heliosheath where the kinematic
approximation of the frozen-in $B$-field
\citep{suessnerney1990}
can be used
and sonic Mach numbers of the plasma are very low, the plasma (also the
electrons) can be treated as incompressible. This may not be a viable
approach near the heliopause where the plasma flow leads to a pile-up of
magnetic field lines
\citep[see e.g.][]{drakeetal2015}.
Here our transport
equations would need corresponding changes.

In view of the electron incompressibility $n_{s_{1}}=n_{s_{2}}$ this leads
to the following simplified relation
\begin{equation}
\label{eq24}
\Delta V_{s_{1}}/\Delta V_{s_{2}}=U_{s_{2}}/U_{s_{1}}
\end{equation}
and consequently leads to the following requirement for the upper
streamline constant $\eta$:
\begin{equation}
\label{eq25}
\exp [\eta]=\bar{U}\bar{\Theta}^{2}\frac{\kappa}{\kappa-3/2}=
\frac{U\Theta^{2}}{U_{0}^{3}}\frac{\kappa}{\kappa-3/2}
\end{equation}
and hence expresses the fact that - in case of an isentropic electron flow -
there exists a concerted or intertwined change of the quantities $U$,
$\kappa$, and $\Theta$ along each streamline in order to fullfill the
thermodynamical relation given in Eq.~(\ref{eq25}).

With the above streamline constant $\exp[\eta]$ one obtains a new
NLTE-behaviour of the electron fluid, namely a concerted variation with the
streamline coordinate $s$ of both kappa function parameters $\Theta$ and
$\kappa$. This, however, has the consequence that the temperatures derived
from the underlying kappa functions are reacting in an unexpected way on
changing values of $\kappa$ when it is taken into account that with a
change in $\kappa$ also a change in $\Theta$ has to occur according to the
following relation
\begin{equation}
\label{eq26}
\frac{U\Theta^{2}}{U_{0}^{3}}\frac{\kappa}{\kappa-3/2}=
\frac{U_{0}\Theta_{0}^{2}}{U_{0}^{3}}\frac{\kappa_{0}}{\kappa_{0}-3/2}
\end{equation}
in other words meaning that in an isentropic electron flow $\Theta $ has to
vary with $\kappa$ according to the following relation
\begin{equation}
\label{eq27}
\Theta^{2}=\frac{\kappa-3/2}{\kappa}\Theta_{0}^{2}\frac{\kappa_{0}}{\kappa_{0}-3/2}\frac{U_{0}}{U}
\end{equation}

From the recent paper by
\citet{fahretal2017},
one can obtain as a starting
value just after passage of the electrons over the termination shock, with
$[\Theta_{0}^{2}/U_{0}^{2}]=0.09\frac{M}{m}$ ($M$ denoting the proton mass, $m$ the electron mass),
$\kappa_{0}=4$ and $U_0=100$~km~s$^{-1}$:
\begin{equation}
\label{eq28}
\lbrack \Theta_{0}^{2}/U_{0}^{2}]\frac{\kappa_{0}}{\kappa_{0}-3/2}=265
\end{equation}

These values are derived in the paper by
\citet{fahretal2017}
with the restriction
that the kappa distributed, shocked electrons represent the right
spectral electron fluxes measured close to the termination shock by
Voyager-1 in the energy range between 40 and 70~keV,
namely 400~electrons/cm$^2$/s/sr/MeV.

\begin{figure}
\resizebox{\hsize}{!}{\includegraphics{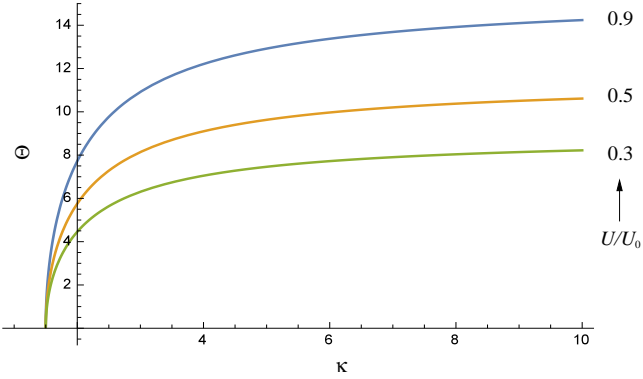}}
\caption{The associated parameter $\Theta$
as a function of the parameter $\kappa$ for the isentropic electron flow
for different values of the electron bulk velocity $U$}
\label{fig1}
\end{figure}

This together with Eq.~(\ref{eq27}) shows the following
variation $\Theta =\Theta(\kappa)$ of the kappa-parameter $\Theta$ with
the kappa-parameter $\kappa$ (Fig.~\ref{fig1}):
\begin{equation}
\label{eq29}
\Theta /U_{0}=\sqrt{265\frac{\kappa -3/2}{\kappa}\frac{U_{0}}{U}}
\end{equation}

In this relation, one may recognize that whenever $\kappa$ approaches the
value $(3/2)$ (i.e.\ extreme suprathermal case), the normalized value of the
associated value $\Theta(\kappa)/U_{0}$ degenerates to $0$ (i.e.\
collapse of the Maxwellian core). For large values of $\kappa$
(i.e.\ Maxwellian case) the associated Maxwellian core temperature will be
higher, the higher is the value $(U/U_{0})$. It is furthermore interesting to
recognize that different, however permitted combinations of parameters
$\kappa$ and the associated $\Theta(\kappa)$, as they can be extracted
from Fig.~\ref{fig1}, belong to different kappa functions which all have the same
entropy, stating the astonishing fact that a complete power law distribution
and the associated Maxwellian with a temperature spread
$\Theta =U_{0}\sqrt{265\frac{U_{0}}{U}}$ have identical entropies.

It is interesting to compare results of kappa-temperatures obtained on the
basis of a constant value of $\Theta$ at varying values of $\kappa$ given by
e.g.\ \citet{heerikhuisenetal2008}
in the form
\begin{equation}
\label{eq30}
T_{\Theta}^{\kappa}(s)=\frac{m\Theta^{2}}{3K}\frac{\kappa}{\kappa-3/2}=
\frac{P^{\kappa}(s)}{nK}
\end{equation}
with corresponding moment-expressions obtained from integrations of the
two-parameter kappa distribution function over the velocity space in the
form:
\begin{equation}
\label{eq31}
T_{\eta}^{\kappa}(s)=\frac{m}{3K}\int_{0}^{v_{\infty}}v^{4}f^{\kappa}(v,\eta (\Theta ,\kappa ))dv
\end{equation}

\begin{figure}
\resizebox{\hsize}{!}{\includegraphics{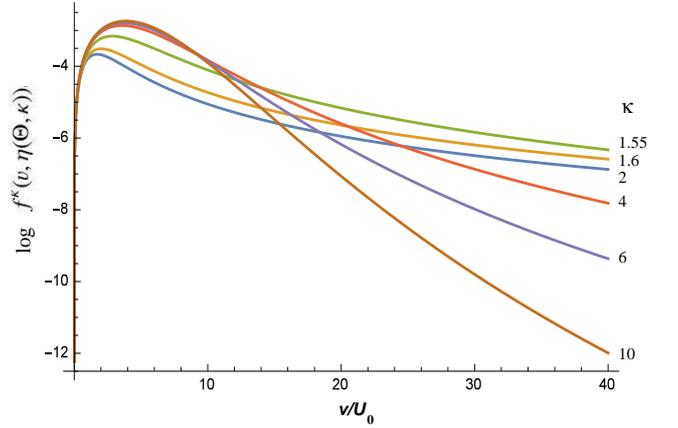}}
\caption{Kappa function $f^{\kappa}(v, \eta(\Theta,\kappa))$ for isentropic electron flows for different values of kappa
as a function of the particle velocity $v$ in units of $U_0$~=~100~km~s$^{-1}$.
One can see that different transitions from a Maxwellian core to a power law distribution are occuring along the heliosheath streamlines at an isentropic electron behaviour.}
\label{fig2}
\end{figure}

\begin{figure}
\resizebox{\hsize}{!}{\includegraphics{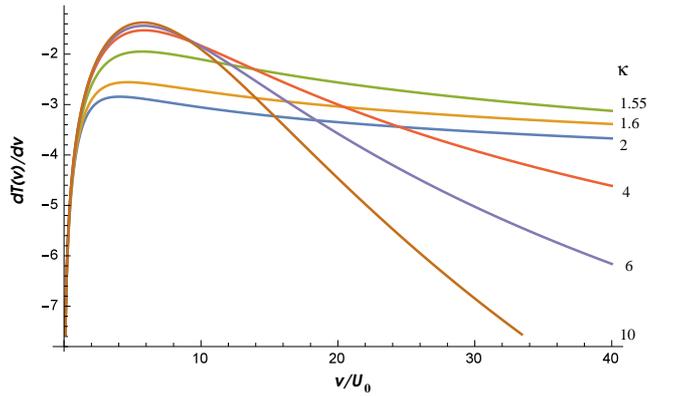}}
\caption{The differential temperature moment of the electrons defined by the integrand of Eq.~(\ref{eq31})
is plotted as a function of the velocity
$v$ in units of $U_0$~=~100~km~s$^{-1}$
for different values of the parameter $\kappa$.
This shows that for different permitted combinations of the parameters $\kappa$ and $\Theta$ the contributions to the effective temperature are due to the higher
electron velocities, the lower are the associated kappa values.}
\label{fig3}
\end{figure}

\begin{figure}
\resizebox{\hsize}{!}{\includegraphics{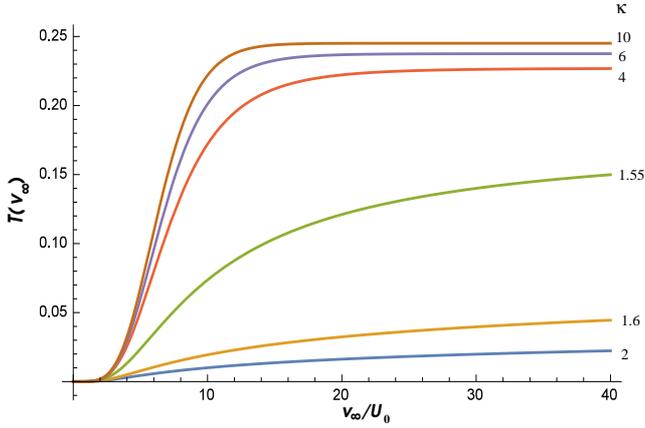}}
\caption{The electron temperature, as a velocity-moment of the distribution function given by Eq.~(\ref{eq31})
as a function of the upper velocity border
$v_{\infty}$ in units of $U_0$~=~100~km~s$^{-1}$,
is shown as a function of the actual kappa value for isentropic
electron flows.}
\label{fig4}
\end{figure}

We have shown in Fig.~\ref{fig2} the distribution function
$f^{\kappa}(v,\eta(\Theta ,\kappa))$ (see Eq.~\ref{eq0})
as a function of the velocity $v$.
In Fig.~\ref{fig3}, we have displayed the differential temperature moment
$dT^{\kappa}/dv$ (i.e.\ the integrand of Eq.~\ref{eq31}).
Finally, in Fig.~\ref{fig4} we show the temperature moment itself as given in Eq.~(\ref{eq31}).

\section{Non-isentropic electron flows and the electron entropy}
In case that the electron flow along the streamlines, contrary to the
assumption in the section before, developes non-isentropically due to the
presence of energy sources or sinks, i.e.\ if the entropy of the electron
fluid changes with the flow line element $s$, then one has to consider the
following more complicate relation along streamlines
\begin{equation}
\label{eq32}
\frac{dS}{ds}=\frac{d}{ds}(\frac{Q}{T})=\frac{d}{dt}(\frac{Q}{T})\frac{1}{U}
=[\frac{\dot{Q}}{T}-\frac{Q}{T^{2}}\dot{T}]\frac{1}{U}=\frac{Q}{T}
[\frac{\dot{Q}}{Q}-\frac{\dot{T}}{T}]\frac{1}{U}
\end{equation}
where $dQ$ according to thermodynamical convention describes energy gains
and losses per unit of the streamline element $s$ in the co-moving volume
$\Delta V$, and $T=T^{\kappa}(s)$ is the local kappa temperature. Looking
at Eq.~(\ref{eq32}), one can already qualitatively say that practically no
change of the entropy has to be faced, if the temporal energy change $\dot{Q}$
compared to the given energy content $Q$ is negligible, i.e.\ the latter
condition for $\frac{\dot{Q}}{Q}=\frac{\dot{\varepsilon}}{\varepsilon}\ll 1$
would lead to a quasi-isentropic behaviour.

In order to check on this condition, we may now look into the
pressure transport equation (Eq.~\ref{eq12}) where one can identify two processes
contributing to the change of the energy density $Q(s)=\varepsilon (s)$,
namely:

1) connected with the process of magnetic cooling (see Section~2,
Eq.~\ref{eq7}) which is given by
\begin{equation}
\label{eq33}
\dot{\varepsilon}_{1}=m \cdot T1=\frac{2}{3}mU\frac{1}{B}\frac{\partial B}{\partial s}
\int dvv^{2}\frac{\partial}{\partial v}(v^{3}f^{\kappa})=\frac{4}{3}U
\frac{1}{B}\frac{\partial B}{\partial s}\frac{3}{4\pi}P^{\kappa}(s)
\end{equation}
and

2) the process of velocity-space diffusion (see Section~2,
Eq.~\ref{eq10}) leading to the following term
\begin{equation}
\label{eq34}
\dot{\varepsilon}_{2}=m \cdot T2=10mD_{0}\int v^{4}f^{\kappa}dv=\frac{3}{4\pi}
10D_{0}P^{\kappa}(s)
\end{equation}

Hence one can express the energy change per time by
$\dot{Q}=\dot{\varepsilon}_{1}+\dot{\varepsilon}_{2}$.

Since we have already derived the solution for the local pressure $P^{\kappa}(s)$,
one thus, based on that, can also obtain the entropy evolution with
the following equation (see Section~2, Eq.~\ref{eq12})
\begin{equation}
\label{eq35}
\frac{dS^{\kappa}}{ds}=\frac{d}{ds}(\frac{Q}{T^{\kappa}})=[\frac{\frac{1}{B}
\frac{\partial B}{\partial s}\frac{U}{\pi}P^{\kappa}(s)+\frac{15}{2\pi}
D_{0}P^{\kappa}(s)}{U\cdot T^{\kappa}(s)}]
\end{equation}
where the actual kappa temperature is given by
\citep[see][]{heerikhuisenetal2008,schereretal2018}
\begin{equation}
\label{eq36}
T^{\kappa}(s)=\frac{m\Theta^{2}}{3K}\frac{\kappa}{\kappa-3/2}=\frac{P^{\kappa}(s)}{nK}
\end{equation}

The kappa-entropy $S^{\kappa}(s)$ can then be calculated with the above
differential equation, using the equation for $P^{\kappa}(s)$ (Eq.~\ref{eq12}) derived
further above in this paper in the form:
\begin{equation}
\label{eq37}
P^{\kappa}(s)=P^{\kappa} (s_{0}) \cdot
\exp [\frac{4}{3} (B(s)-B(s_{0}))+10D_{0}
\int_{s_{0}}^{s}\frac{ds}{U}]
\end{equation}
and hence now obtaining the following differential equation
\begin{equation}
\begin{split}
\label{eq38}
\frac{dS^{\kappa }}{ds} &= [nK\frac{\frac{1}{B}\frac{\partial B}{\partial s}
\frac{U}{\pi}P^{\kappa}(s)+\frac{15}{2\pi}D_{0}P^{\kappa}(s)}{U\cdot
P^{\kappa}(s)}] \\
&= [\frac{nK}{\pi}(\frac{1}{B}\frac{\partial B}{\partial s}+
\frac{15}{2}\frac{D_{0}}{U}]
\end{split}
\end{equation}
which can be integrated to yield
\begin{equation}
\begin{split}
\label{eq39}
S^{\kappa}(s) &= S^{\kappa}(s_{0})+\frac{nK}{\pi }\int_{s_{0}}^{s}(\frac{1}{B}
\frac{\partial B}{\partial s}+\frac{15}{2}\frac{D_{0}}{U})ds \\
&= S^{\kappa}(s_{0})
+\frac{nK}{\pi}[\ln (\frac{B(s)}{B(s_{0})})+\frac{15D_{0}}{2}
\int_{s_{0}}^{s}\frac{ds}{U}]
\end{split}
\end{equation}

Now it is again interesting to see that also the entropy $S^{\kappa}(s)$,
as the pressure $P^{\kappa}(s)$, depends directly obviously
neither on $\kappa$ nor on $\Theta$, since it turns out according
to the above relation that $\partial S^{\kappa}(s)/\partial \kappa
=\partial S^{\kappa}(s)/\partial \Theta =0$.

On the one hand we have found that isentropic electron flows require the
combined change of the local kappa parameters $\kappa$ and $\Theta$, on
the other hand the entropy itself, controlling the isentropic behaviour,
does not directly depend on the kappa parameters themselves. Looking finally
on the amount of entropy changes that are to be expected according to the
formula above, it turns out, in view of the quantities entering Eq.~(\ref{eq39}),
that the resulting entropy change, at least on the frontal side
of the heliosheath (upwind side), will be fairly small.

\section{Conclusions}
In this paper, we have derived a kinetic transport equation describing the
evolution of the heliosheath electron distribution along the flowlines of
the plasma bulk flow downstream of the solar wind termination shock. From
this kinetic transport equation one can proceed to moment transport
equations. Here, we especially develop and mathematically treat the pressure
transport equation which describes the evolution of the electron pressure
along the plasma flow lines. It turns out that on the basis of the processes
taken into account which influence electrons at their passage along the
flowlines, namely convective phase-space changes, magnetic cooling processes
and wave-particle diffusion, an analytic solution for the electron pressure
as function of the streamline coordinate can be obtained. Introducing most
probable parameter values for the relevant processes, it turns out
that, at least on the upwind side of the heliosheath, the electron pressure
is essentially constant along the flow lines, only varying from streamline
to streamline (i.e.\ perpendicular to the streamlines) due to different
boundary conditions at the flowline origin at the termination shock where,
due to different injection conditions, the injected electrons lead to
different initial pressures. This means, nevertheless,
that the flowlines serve more or less as electron isobars.

It is then studied by us, how isentropic electron flows with $dS_{e}/ds=0$
should behave, and it is shown, that, under isentropic electron flow
conditions, the two parameters of the electron kappa function $\kappa(s)$
and $\Theta(s)$ should undergo concerted changes with the streamline
coordinate $s$, i.e.\ in the form $\Theta(s)=\Theta (\kappa(s))$. We show
that a streamline parameter $\eta$ exists which describes these concerted
changes along streamlines in the form $d\eta/ds=d[\eta(s,\kappa,\Theta)]/ds=0$.
Thus, for isentropic electron flows, the two parameters $\kappa(s)$
and $\Theta(s)$ of the electron kappa-function along the streamline
undergo concerted variations as described by Eq.~(\ref{eq29}).

This means that the entropy of a kappa function $S(\kappa,\Theta)$ for
different values $\kappa$ and $\Theta$ leads to the same value
$S=S(\kappa(\eta),\Theta(\eta))$, if the two parameters are intertwined according to Eq.~(\ref{eq29}).
This also allows to conclude that for $1\ll \kappa$
one finds entropies identical to the associated Boltzmannian or 
Maxwellian quasi-LTE entropies $S(T_{\max})$ connected with a temperature of 
\begin{equation}
\label{eq40}
T_{\max}=\frac{3}{2}m\Theta^{2}(\kappa \rightarrow \infty)=\frac{3}{2}
m\cdot (265U_{0}/U).
\end{equation}

On the other hand, we have studied non-isentropic electron flows, including
energy sources and sinks operating on the convected electrons, like
magnetic cooling or energy-diffusion processes. This allows to derive
an entropy transport equation of the form given by Eq.~(\ref{eq39}) and to obtain
an analytic solution for the electron entropy $S(s)$ as function of the
streamline coordinate. Under the conditions prevailing in the heliosheath,
i.e.\ subsonic flow, incompressible behaviour, etc., one finds that the
variation of the electron entropy along flowlines is only weakly pronounced.
The electron flows along streamlines in the heliosheath can thus be essentially characterized
as incompressible, isobaric, isothermal and isentropic flows.

\section*{Acknowledgements}
We hereby acknowledge the participation of H.\ J.\ Fahr at the first ISSI meeting on Kappa-distributions at ISSI, Bern, Feb. 12-16, 2018, and many soliciting discussions connected with it.

\bibliographystyle{mnras}
\bibliography{ref}

\bsp	
\label{lastpage}
\end{document}